\documentclass[twocolumn,showpacs,aps,superscriptaddress]{revtex4-1}
\usepackage{amsmath}
\usepackage{amsfonts}
\usepackage{amssymb}
\usepackage{dcolumn}
\usepackage[dvips]{graphicx}
\usepackage{epsfig}
\usepackage{bm}

\begin{document}

\title{Fragility of superposition states evaluated by the Loschmidt Echo}

\author{Denise Bendersky}
\affiliation{Instituto de F\'{i}sica Enrique Gaviola (IFEG), CONICET-UNC and Facultad
de Matem\'{a}tica, Astronom\'{i}a y F\'{i}sica, Universidad Nacional
de C\'{o}rdoba, 5000, C\'{o}rdoba, Argentina}
\affiliation{Instituto de Astronom\'{\i}a y F\'{\i}sica del Espacio (IAFE), CONICET-UBA}
\author{Pablo R. Zangara}
\affiliation{Instituto de F\'{i}sica Enrique Gaviola (IFEG), CONICET-UNC and Facultad
de Matem\'{a}tica, Astronom\'{i}a y F\'{i}sica, Universidad Nacional
de C\'{o}rdoba, 5000, C\'{o}rdoba, Argentina}
\author{Horacio M. Pastawski}
\email{horacio@famaf.unc.edu.ar}
\affiliation{Instituto de F\'{i}sica Enrique Gaviola (IFEG), CONICET-UNC and Facultad
de Matem\'{a}tica, Astronom\'{i}a y F\'{i}sica, Universidad Nacional
de C\'{o}rdoba, 5000, C\'{o}rdoba, Argentina}

\begin{abstract}
We consider the degradation of the dynamics of a Gaussian wave packet in a harmonic oscillator under the presence of an environment. This last is given by a single non-degenerate two level system. We analyze how the binary degree of freedom perturbs the free evolution of the wave packet producing decoherence, which is quantified by the Loschmidt echo. This magnitude measures the reversibility of a perturbed quantum evolution. In particular, we use it here to study the relative ``fragility'' of coherent superpositions (cat states) with respect to incoherent ones. This fragility or sensitivity turns out to increase exponentially with the energy separation of the two
components of the superposition.
\end{abstract}

\maketitle

\section{Introduction}

\label{sec_intro}

Control and manipulation of coherent quantum systems is a major task for both nanotechnology \cite{vandersypen_RMP,flatte07} and fundamental physics \cite{RevModPhysBloch08}. Specifically, quantum information processing (QIP) operates with superposition states, which constitute the heart of quantum weirdness \cite{vicenzo}. Since these states have an intrinsic nonlocal nature, their characterization becomes a nontrivial problem \cite{horodecki} . Moreover, the prediction and control of their time evolution are restricted by the unavoidable interactions with an environment that degrades the unitarity of quantum dynamics \cite{ZurekRevMod}. This process, called \textit{decoherence}, involves the progressive and smooth destruction of the quantum interferences \cite{Imry1990}.

Among the general expectations within the field quantum open systems \cite{ZurekNature}, is the claim that the more non-local and complex a superposition state is, the more fragile it becomes under the effects of decoherence. If indeed general, this might preclude scalability of QIP implementations. The magnitude of such fragility seems to be intimately related to the number of correlated qubits and the way in which they evolve. In particular, nuclear magnetic resonance (NMR) experiments with \textit{large} arrays of interacting spins \cite{attenuationNMR98,Suter_PRL2004,SanchezLevsteinAcostaChattahPRA09} have shown that these can exhibit an intrinsically unstable dynamics. Inspired by the NMR experiments, the Loschmidt echo (LE) \cite{Gorin,jacquod,scholarpedia} arises as a natural way to quantify fragility. The LE is defined in terms of the revival that occurs when a slightly imperfect time-reversal procedure is applied to a quantum evolution. Such imperfection accounts for the presence of uncontrolled degrees of freedom, which play the role of an environment \cite{zurekpolonica,Jacquod-prl2006,pablo2012}. Quite remarkably, it has been proved that even in the presence of simple perturbations, chaotic systems can become their own environment \cite{JalabertPastawski}. Furthermore, in such classically chaotic systems, dynamics leads to highly nonlocal superpositions that have already been related to the formation of sub-Planck-scale structures associated with a boost of decoherence \cite{Jacquod2002,Zurek2002}.

The standard theoretical strategy to address decoherence and dissipation relies on defining a simple system $\mathcal{S}$ that interacts with a large and complex environment $\mathcal{E}$ \cite{bookBreuer2002,bookSchlosshauer07}. While the former has a few degrees of freedom, the latter typically has a dense spectrum, at least within the experimental time scales shorter than the Heisenberg time, in which mesoscopic echoes would show up \cite{Altshuler94,mesoECO-theory}. Within this framework, the spin-boson model (SBM) turns out to be one of the most employed paradigms \cite{Legget_chakravarty_RevModPhys}. This corresponds to a single two-level system (TLS) $\mathcal{S}$ interacting with a large reservoir $\mathcal{E}$ of bosonic field modes, i.e., a spin 1/2 coupled to an environment of harmonic oscillators (HO). This model has also found wide application in the fields of chemical and biological physics, providing a rationale for the electron-transfer process. There, the role of the spin is played by a charge that can fluctuate between two reaction centers \cite{Marcus_RevModPhys, bookNitzan}.

In this article, we switch the spin and boson roles as $\mathcal{S}$ and $\mathcal{E}$, using a HO as $\mathcal{S}$ and a single TLS as $\mathcal{E}$. This crucially different point of view seeks to assess how states with controllable complexity are degraded by a simple $\mathcal{E}$. Specifically, only when this binary degree of freedom flips its state is the mixing among the states of $\mathcal{S}$ enabled. This approach allows us to test the above assumption about the fragility of specific nonlocal superpositions, or \textit{cat} states. Here, these highly nonlocal superpositions are not obtained as dynamically prepared initial states on a chaotic system \cite{Jacquod2002,Zurek2002} but are built as specific initial states of the HO. In the particular case analyzed here, \textit{coherent} superpositions of two semiclassical states associated with different energies show an enhanced fragility with respect to the \textit{incoherent} superpositions.

This paper is organized as follows. In Sec. \ref{sec_SBM} we describe our version of the SBM and summarize its theoretical background. In Sec. \ref{sec_InitialStates} we define the semiclassical and the superposition states built with them. The incoherent superposition is described in detail in the Appendix. In Sec. \ref{sec_LE} we describe how the LE is evaluated for mixed states without resorting to the evaluation of the density matrix, i.e. just from the wave function in the Fock space. This
LE evaluation is explained in some detail in the Appendix. In Sec. \ref{sec_LZ} we show that the dynamics can be analyzed in terms of the Landau-Zener (LZ) theory \cite{landau,zener,LZ_review}. In Sec. \ref{sub_fragility} we quantify the fragility of different initial states. Finally, a brief discussion of the results and conclusions is presented in Sec. \ref{sec_conclusions}.

\section{The spin-boson model}

\label{sec_SBM}

\begin{figure*}
    \centering
     \includegraphics[width=0.70\textwidth]{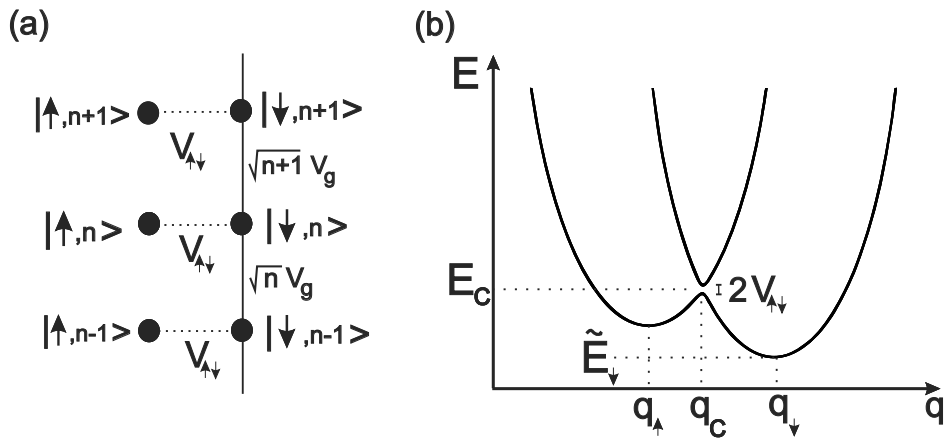}\\
      \caption{(a) Fock space representation
of $\hat{H}=\hat{H}_{\mathcal{S}}$ $+\hat{H}_{\mathcal{E}}+\hat{V}_{\mathcal{SE}}$. Vertical hoppings enable the mixing processes among the HO eigenbasis. (b) Semiclassical representation of $\hat{H}$ in terms of the HO's canonical coordinate $\hat{q}$ . The energy is expressed in units of $\hbar \omega_0$, and the space coordinate is in units of $\sqrt{\hbar/\left(m\omega^{}_0\right)}$. Each parabola corresponds to a different spin state while the spin-flip process $V_{\uparrow \downarrow }$ produces an anticrossing (energy gap).}
    \label{fig_model}%
\end{figure*}

We propose that the system $\mathcal{S}$ under consideration is a HO which is coupled to a TLS, which constitutes the environment $\mathcal{E}$. This TLS enables mixing and produces decoherence, causing a loss of control over $\mathcal{S}$. The total Hamiltonian which contains all these physical processes is
\begin{equation}
\hat{H}=\hat{H}_{\mathcal{S}}\otimes \hat{I}_{\mathcal{E}}+\hat{I}_{\mathcal{%
S}}\otimes \hat{H}_{\mathcal{E}}+\hat{V}_{\mathcal{SE}},  \label{eq htot}
\end{equation}%
where the first term represents $\mathcal{S}$:

\begin{equation}
\hat{H}_{\mathcal{S}}=\hbar \omega _{0}\left( \hat{b}^{+}\hat{b}+\frac{1}{2}%
\right) ,  \label{eq hsys}
\end{equation}%
with $\hat{b}^{+}$ being and $\hat{b}$ the bosonic raising and the lowering operators respectively. The second term in Eq.\ref{eq htot} represents the TLS, which corresponds to $\mathcal{E}$,

\begin{equation}
\hat{H}_{\mathcal{E}}=E_{\uparrow }\hat{c}_{\uparrow }^{+}\hat{c}_{\uparrow
}^{{}}+E_{\downarrow }\hat{c}_{\downarrow }^{+}\hat{c}_{\downarrow
}^{{}}+V_{\uparrow \downarrow }\left( \hat{c}_{\uparrow }^{+}\hat{c}%
_{\downarrow }^{{}}+\hat{c}_{\downarrow }^{+}\hat{c}_{\uparrow }^{{}}\right)
,  \label{eq henv}
\end{equation}%
where $\hat{c}_{s}^{+}$ and $\hat{c}_{s}^{{}}$ ($s\in \left\{ \uparrow ,\downarrow \right\} $) are the creation and destruction operators for fermions. Within an electron-transfer model, $E_{\uparrow }$ and $E_{\downarrow }$ are the nondegenerate electron's energies at states $\uparrow $ and $\downarrow $, respectively.\ Since the Wigner-Jordan transformation allows for a precise correspondence between spinless fermions and spin states, the hopping amplitude $V_{\uparrow \downarrow }~$ (i.e., the electron's tunneling between the centers $\uparrow $ and $\downarrow $) also describes a spin-flip process. As the interaction between $\mathcal{S}$ and $\mathcal{E}$ we adopt the standard linear electron-phonon interaction, used to describe the Franck-Condon effect and electron transfer processes \cite{polianilina}:

\begin{equation}
\hat{V}_{\mathcal{SE}}=-V_{g}\left( \hat{b}^{+}+\hat{b}\right) \hat{c}
_{\downarrow }^{+}\hat{c}_{\downarrow }^{{}},  \label{eq vint}
\end{equation}
where $V_{g}$ gives the scale for the $\mathcal{S}$-$\mathcal{E}$ interaction which is strong enough to regard the $\mathcal{S}$ spectrum as quasicontinuous ($V_{g}\gg \hbar \omega^{} _{0}$). In this model, it is clear that the state of the HO $\mathcal{S}$ and its dynamics depend on the spin state. Indeed, Eq.\ref{eq vint} implies an explicit displacement in the harmonic potential which is evidenced when the total Hamiltonian $\hat{H}$ is written in terms of the two canonical coordinates $\hat{p}$ and $\hat{q}$:
\begin{equation}
\hat{H}_{\mathcal{S}}+\hat{V}_{\mathcal{SE}}=\frac{\hat{p}^{2}}{2m}+\frac{1}{%
2}m\omega _{0}^{2}\hat{q}^{2}-\sqrt{\frac{2m\omega _{0}}{\hbar }}V_{g}\hat{q}%
\hat{c}_{\downarrow }^{+}\hat{c}_{\downarrow }^{{}},  \label{eq Hpq}
\end{equation}
where $\hat{q}=\sqrt{\frac{\hbar }{2m\omega _{0}}}\left( \hat{b}^{+}+\hat{b}\right) $ and $\hat{p}=i\sqrt{\frac{\hbar }{2m\omega _{0}}}\left(\hat{b}^{+}-\hat{b}\right)$. In terms of the canonical coordinate $\hat{q}$, the Hamiltonian given in Eq. \ref{eq Hpq} makes the potential surface shifted with respect to the one corresponding to the usual HO. The model is schematized in Fig. \ref{fig_model}, where in Fig. \ref{fig_model}(a) the interactions are represented in the Fock space. In \ref{fig_model}(b) we show a semiclassical representation of the perturbed potential, where we can notice the energy gap of width $2V_{\uparrow \downarrow }$ induced by $\hat{H}_{\mathcal{E}}$, with parameters
\begin{eqnarray}
q_{C} &=&\sqrt{\frac{\hbar }{2m\omega _{0}}}\frac{E_{\downarrow }}{V_{g}},
\label{eq_cross} \\
E_{C} &=&\frac{\hbar \omega _{0}}{4}\left( \frac{E_{\downarrow }}{V_{g}}\right) ^{2}.
\end{eqnarray}

In order to simplify the analysis we define an energy reference which shifts the perturbed harmonic potential, producing $E_{\uparrow }-\tilde{E}_{\downarrow }=0$ with $\tilde{E}_{\downarrow }=E_{\downarrow}-V_{g}^{2}/\hbar \omega _{0}$. This means that the parabolas in Fig. \ref{fig_model}(b) are symmetrical with respect to the crossing point $q_{C}$. Thus, similar energies in the HO are mixed up by the dynamics of the environment. In fact, the HO energy density $1/\hbar \omega _{0}$ constitutes the quasicontinuous spectrum that can be easily mixed up by the dynamics of the
TLS. This would require that $V_{\uparrow \downarrow }\gg \hbar \omega _{0}$.

\section{Initial states}

\label{sec_InitialStates}

\begin{figure*}
    \centering
     \includegraphics[width=0.9\textwidth]{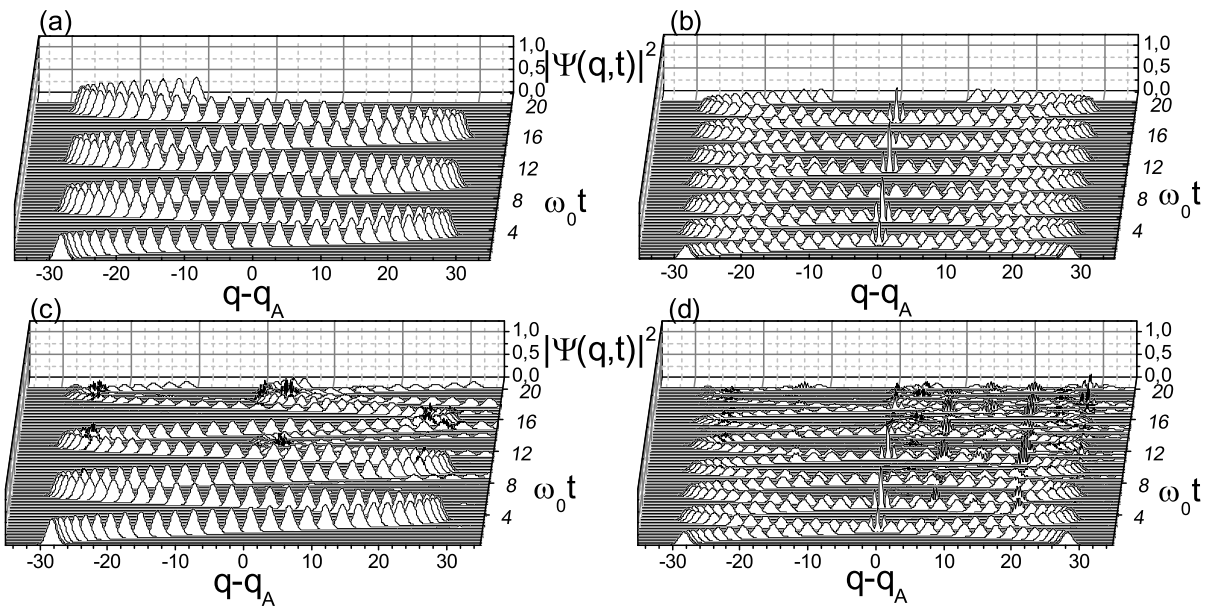}\\
      \caption{Time evolution of the probability distribution as a function of the space coordinate, which is expressed in units of $\sqrt{\hbar/\left(m\omega^{}_0\right)}$. (a) Unperturbed case for an initial semiclassical state with $\bar{E}=400\hbar \protect\omega^{} _{0}$. (b) Unperturbed case for an initial cat state with the same energy. The perturbed dynamics for the same initial states are shown in (c) and (d), respectively. }
    \label{fig_evolution}
\end{figure*}

Within the tight-binding representation of the Fock space shown in Fig. \ref{fig_model}(a), any wave function for the whole $\mathcal{S+E}$ can be written as
\begin{equation}
\left\vert \Phi \right\rangle =\sum_{k=\uparrow ,\downarrow
}\sum_{n=0}^{\infty }c_{k,n}\left\vert k,n\right\rangle ,
\label{eq general_wave}
\end{equation}%
where the probability amplitudes $c_{k,n}$ have the $k$ index that labels the spin states $\left\vert \uparrow \right\rangle$, and $\left\vert\downarrow \right\rangle$ and the $n$ index that labels the HO eigenstates. We consider three different initial states, all restricted to the TLS state $\left\vert \uparrow \right\rangle$. The first one, a Gaussian wave packet usually called a coherent state, is

\begin{equation}
\left\vert \alpha \right\rangle =e^{-\frac{\left\vert \alpha \right\vert ^{2}}{2}}\sum\limits_{n=0}^{\infty }\frac{\alpha ^{n}}{\sqrt{n!}}\left\vert
\uparrow ,n\right\rangle .  \label{eq coherent}
\end{equation}
In order to avoid confusion, from now on we will refer to such states as \textit{semiclassical} since they exhibit a minimum uncertainty. Its energy is given by $E_{\alpha }=\hbar \omega _{0}\left( \left\vert \alpha\right\vert ^{2}+1/2\right)$, and its evolution under the unperturbed Hamiltonian $\hat{H}_{\mathcal{S}}$ is a trivial semiclassical oscillation. The second case for the initial state is a non-local or cat state, i.e., a coherent superposition of two Gaussians,

\begin{equation}
\left\vert \Psi _{0}^{cat}\right\rangle =\frac{\left\vert \alpha_{1}\right\rangle +\left\vert
\alpha_{2}\right\rangle }{\sqrt{2}},
\label{eq cat}
\end{equation}
where $\left\vert \alpha _{r}\right\rangle $ is a semiclassical state as in Eq. \ref{eq coherent}, associated with a complex number $\alpha _{r}$. Notice that we consider $\alpha _{1}$ and $\alpha _{2}$ with different signs, but they may not have equal modulus. Since the energy is proportional to $\vert \alpha_{r} \vert ^2$, the cat states introduced here can involve the superposition of two Gaussians with different energies as in Ref. \cite{deutsch}. The third case considered is an incoherent superposition, which is written as
\begin{equation}
\left\vert \Psi _{0}^{inc}\right\rangle =\lim_{N\rightarrow \infty
}\sum\limits_{j=1}^{N}\frac{e^{i\theta _{j}}\left\vert \alpha
_{1}\right\rangle +e^{i\phi _{j}}\left\vert \alpha
_{2}\right\rangle }{\sqrt{\Delta }}, \label{eq incoh}
\end{equation}
where $\theta _{j}$ and $\phi _{j}$ are random variables uniformly distributed in $[0,2\pi )$ and the normalization factor $\Delta^{2}=2N^{2}(3+\exp [-\left\vert \alpha _{1}-\alpha _{2}\right\vert ^{2}])$ is explicitly computed in the Appendix. The random phase relation of this state leads to two local probability distribution functions with the same statistical weights. It is the analog to the random superpositions employed in spin systems to simulate high-temperature states \cite{Alv-parallelism,Pablo_LSantos}. Notice that in these three cases we restricted the initial state to a definite spin projection $\left\vert\uparrow \right\rangle$.

In Fig. \ref{fig_evolution} we illustrate the evolution in space for an initial semiclassical state Fig. \ref{fig_evolution}(a) and an initial cat state Fig. \ref{fig_evolution}(b) under the action of the unperturbed Hamiltonian $\hat{H}_{\mathcal{S}}$, while in Figs. \ref{fig_evolution}(c) and \ref{fig_evolution}(d) we show the evolution of the same initial states under the action of the total Hamiltonian $\hat{H}$. In the last two cases, it can be noticed how
the oscillatory dynamics is perturbed by successive passages through the avoided crossing region around $q=q_{C}$. Notice that the amplitude of the semiclassical oscillations remains unaffected at least during several cycles, which means that there is no considerable loss of energy due to the
interaction. The splittings of the wave packet trajectories will be analyzed within the LZ theory in Sec. \ref{sec_LZ}. The coherent and incoherent superpositions defined here constitute the trial states which will be employed to evaluate the \textit{fragility} in Sec. \ref{sub_fragility}.

\section{Loschmidt Echo}

\label{sec_LE}

We employ the LE as a decoherence quantifier and our specific purpose relies on analyzing the \textit{fragility} of the states introduced in Sec. \ref{sec_InitialStates}. In fact, the LE measures the sensitivity of a quantum evolution to non controlled perturbations \cite{Gorin,jacquod,scholarpedia}.
It relies on a time-reversal procedure within $\mathcal{S}$ degrees of freedom, which filters out $\mathcal{S}$ dynamics and allows us to address the degradation induced by the $\mathcal{E}$ degrees of freedom. For an initial state $\left\vert \Psi _{0}\right\rangle $ that describes the whole $\mathcal{S+E}$, the standard LE formula is \cite{JalabertPastawski}

\begin{equation}
M(t)=\left\vert \left\langle \Psi _{0}\right\vert \exp \left\{ \frac{\mathrm{%
i}}{\hbar }\left( \hat{H}_{\mathcal{S}}+\hat{\Sigma}\right) t\right\} \exp
\left\{ -\frac{\mathrm{i}}{\hbar }\hat{H}_{\mathcal{S}}t\right\} \left\vert
\Psi _{0}\right\rangle \right\vert ^{2}.  \label{eq_LEformula1}
\end{equation}%
Here, the perturbation operator $\hat{\Sigma}$ represents $\hat{H}_{\mathcal{E}}+\hat{V}_{\mathcal{SE}}$ as defined in Sec. \ref{sec_SBM}. The state $\left\vert \Psi _{0}\right\rangle$ evolves forward in time with $\hat{H}_{\mathcal{S}}$, i.e., without interacting with $\mathcal{E}$, which remains frozen. This evolution can be written in an analytically closed form. At time $t$, an imperfect time-reversal procedure is applied within $\mathcal{S}$ that nevertheless is unable to decouple $\mathcal{S}$ from $\mathcal{E}$. Further evolution during a symmetric backward period occurs under the full Hamiltonian. Thus, the uncontrolled degrees of freedom  lead to the degradation of the overlap between the initial and the time-reversed wave functions.

The LE as defined in Eq.\ref{eq_LEformula1} is not appropriate since it implies a raw overlap of both the $\mathcal{S}$ and the $\mathcal{E}$ components of two wave functions. As discussed above, we are specifically interested in evaluating how the HO ($\mathcal{S}$) is perturbed by the
binary degree of freedom ($\mathcal{E}$). Thus, it is necessary to perform a partial trace over the $\mathcal{E}$ degrees of freedom \cite{Jacquod-prl2006,pablo2012}. Let us define two states of the whole $\mathcal{S+E}$ from which the LE is evaluated, in the explicit form of Eq.\ref{eq general_wave}:
\begin{eqnarray}
\left\vert \Psi (t)\right\rangle &=&e^{-\frac{i}{\hbar }\hat{H}_{\mathcal{S}}t}\left\vert \Psi _{0}\right\rangle =\sum_{n=0}^{\infty
}c_{\uparrow ,n}^{{}}(t)\left\vert \uparrow ,n\right\rangle ,  \nonumber \\
\left\vert \Phi (t)\right\rangle &=&e^{-\frac{i}{\hbar }(\hat{H}_{\mathcal{S}}+\hat{\Sigma})t}\left\vert \Psi _{0}\right\rangle
=\sum_{n=0}^{\infty }\sum_{k=\uparrow ,\downarrow }d_{k,n}^{{}}(t)\left\vert
k,n\right\rangle .  \label{eq expansion}
\end{eqnarray}

Now we trace over the $\mathcal{E}$ degrees of freedom to build the reduced density operators,
\begin{eqnarray*}
\sigma _{\mathcal{S}}^{r} &\equiv &\mathrm{Tr}_{\mathcal{E}}\left(
\left\vert \Psi \right\rangle \left\langle \Psi \right\vert \right)
=\sum_{m,n=0}^{\infty }\left[ c_{\uparrow ,n}^{{}}(t)c_{\uparrow ,m}^{\ast
}(t)\right] \left\vert n\right\rangle \left\langle m\right\vert , \\
\sigma _{\mathcal{S+E}}^{r} &\equiv &\mathrm{Tr}_{\mathcal{E}}\left(
\left\vert \Phi \right\rangle \left\langle \Phi \right\vert \right)
=\sum_{m,n=0}^{\infty }\left[ \sum_{k=\uparrow ,\downarrow
}d_{k,n}^{{}}(t)d_{k,m}^{\ast }(t)\right] \left\vert n\right\rangle
\left\langle m\right\vert ,
\end{eqnarray*}%
where the spin index is no longer present in the bra-ket basis. The LE is now defined as the overlap of these reduced states and can be explicitly written as
\begin{align}
M(t) &=\mathrm{Tr}\left\{ \left( \sigma _{\mathcal{S}}^{r}\right) ^{\dag
}\sigma _{\mathcal{S+E}}^{r}\right\}  \label{eq_LEformula2} \\
&=\sum_{m,n=0}^{\infty }\left[ d_{\uparrow ,m}^{{}}(t)d_{\uparrow ,n}^{\ast
}(t)+d_{\downarrow ,m}^{{}}(t)d_{\downarrow ,n}^{\ast }(t)\right]  \nonumber \\
& \qquad \qquad \times \left[
c_{\uparrow ,n}^{{}}(t)c_{\uparrow ,m}^{\ast }(t)\right] .
\label{eq_LEformula3}
\end{align}%

Here, one can notice that in spite of the \textit{formal} use of density matrices, an actual LE computation can avoid any matrix manipulation at all. Indeed, the equality in Eq.\ref{eq_LEformula2} gives a direct recipe to evaluate the LE from specific components of the wave function $\left\vert
\Phi \right\rangle $ in the Fock space, explicitly given in Eq.\ref{eq expansion}. Even though the LE in Eq.\ref{eq_LEformula3} is written in terms of products of complex amplitudes, we stress that by construction, it is in fact a real and positive quantity.

Notice that if the initial state is a superposition (both cat or incoherent) the linearity of the evolution operators can be employed to evaluate the probability amplitudes as a sum of two contributions. This is explicitly used in the LE computation for the incoherent superposition shown in the Appendix. There, the phase averaging is performed and a particular version of Eq. \ref{eq_LEformula3} is derived. Additionally, a naive version of the LE is obtained by averaging independent realizations of the echo procedure for each of the single semiclassical states, $\left\vert \alpha _{1}\right\rangle $ and $\left\vert \alpha _{2}\right\rangle $.

\section{Results}

\label{sec_results}

\subsection{The Landau-Zener picture}

\label{sec_LZ}

\begin{figure}
    \centering
     \includegraphics[width=0.5\textwidth]{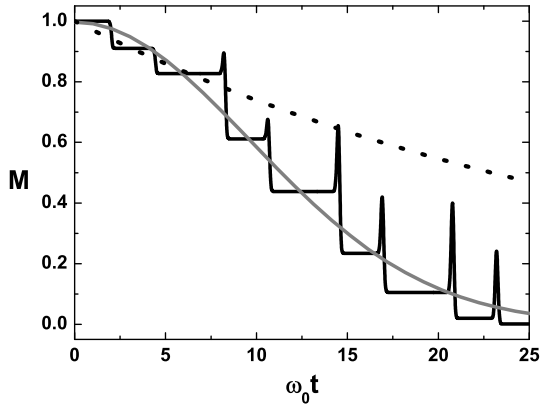}\\
      \caption{Loschmidt echo (black curve) of an initial semiclassical state with energy $E_{0}=200\hbar \protect\omega^{}_{0}$. The dotted curve represents a Markovian approximation by means of an exponential decay, and the gray curve represents a Gaussian fitting to the decay (see text). The parameters used in the LE dynamics are $E_{\uparrow }=0$, $E_{\downarrow }=100\hbar \protect\omega^{}_{0}$, $V_{g}=10\hbar \protect\omega^{} _{0}$, $V_{\uparrow \downarrow }=2\hbar \protect\omega^{} _{0}$. }
    \label{fig LE1}
\end{figure}

Since the initial states given by Eqs.\ref{eq coherent},\ref{eq cat} and \ref{eq incoh} are explicitly defined with the spin state $\left\vert \uparrow \right\rangle $, then the transitions among the HO eigenstates are forbidden unless the spin flips to $\left\vert\downarrow\right\rangle$. This is explicitly shown by the tight-binding representation in Fig.\ref{fig_model}(a) and the harmonic potential in Fig.\ref{fig_model}(b). As already pointed above, the term $\hat{H}_{\mathcal{E}}$\ in the total Hamiltonian $\hat{H}$\ produces an avoided level crossing. A semiclassical wave packet evolving in the presence of the harmonic potential does not degrade unless it goes through such an energy gap. In fact, decoherence processes induced by $\hat{V}_{\mathcal{SE}}$ are enabled only if a passage takes place, which means that they are restricted to a specific region in space and time.

In all cases considered here, we fix the parameters of the model in such a way that the potential energy parabolas are at the same height. We choose parameters satisfying the assumptions discussed above: $V_{g}=10\hbar \omega^{} _{0}$. Thus, with $E_{\uparrow }\equiv 0$ one gets $E_{\downarrow }=100\hbar \omega^{} _{0}$ and $E_{C}=25\hbar \omega^{} _{0}$, which is always much smaller than the energy $E_{0}$ of the initial state. Also, we choose $\alpha \in \mathbb{R}$, so that the initial wave packet velocity is zero. In Fig. \ref{fig LE1} we show the LE decay for an initial semiclassical state, which consistently evidences a discrete set of steps that are associated with each passage through the avoided crossing.

\begin{figure}
    \centering
     \includegraphics[width=0.5\textwidth]{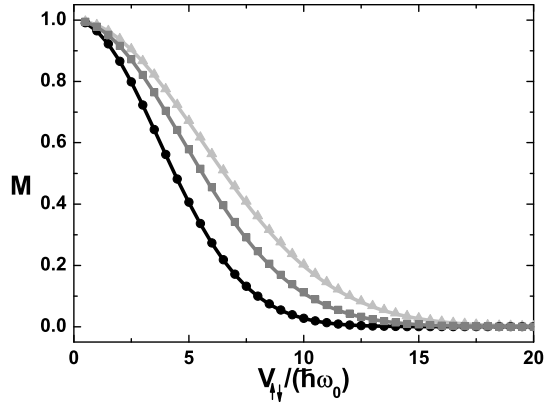}\\
      \caption{Comparison between the LE evaluated in the first step and the LZ probability (solid curves) as a function of $V_{\uparrow \downarrow }$ for different wave-packet energies. From top to bottom, light gray triangles correspond to $\bar{E}=400.5\hbar \protect\omega^{} _{0}$, gray squares correspond to $\bar{E}=225.5\hbar \protect\omega^{} _{0}$ and black circles correspond to $\bar{E}=100.5\hbar \protect\omega^{} _{0}$.}
    \label{fig_probLZ}
\end{figure}

In order to quantitatively analyze the LE decay in our quasicontinuous system we follow Marcus \cite{Marcus_RevModPhys} by identifying it with the LZ problem. This involves the evaluation of the transition probability within a two-level system under the action of a time-dependent bias. We have already shown in Fig.\ref{fig_evolution} that the\ wave packet splits every time that it crosses the gap, which occurs at $q=q_{C}$, given by Eq.\ref{eq_cross}. At time $t=t_{C}$ the wave packet goes through the region of avoided crossing with an approximately constant velocity $\dot{q}_{C}$. If $\mathcal{S}$ is in the spin state $\left\vert \uparrow \right\rangle $, the LZ asymptotic probability to remain in the spin state $\left\vert \uparrow \right\rangle $ is given by

\begin{equation}
P_{\uparrow \uparrow }(t\rightarrow \infty )=\exp \left\{ -\frac{2\pi }{\hbar }\frac{\left\vert V_{\uparrow \downarrow }\right\vert ^{2}}{\left\vert
\frac{d}{dt}\left( \epsilon _{\uparrow }-\epsilon _{\downarrow }\right)
\right\vert }\right\} ,  \label{eq_LZ}
\end{equation}
where $\epsilon _{\uparrow }(q)$ and $\epsilon _{\downarrow }(q)$ are the potential energies described by linear approximations for their $q$ dependence, i.e., $\epsilon _{\uparrow }\approx m\omega
_{0}^{2}q_{C}^{2}/2+m\omega _{0}^{2}q_{C}(q-q_{C})$ and $\epsilon
_{\downarrow }\approx m\omega _{0}^{2}q_{C}^{2}/2-m\omega_{0}^{2}q_{C}(q-q_{C})$. In turn, this becomes a dependence on time $t$ if one assumes  that near the crossing point the wave packet behaves linearly as $q\equiv \left\langle q(t)\right\rangle \simeq $ $q_{C}+\dot{q}_{C}(t-t_{C})$. Thus, at least in a single passage, we can map the conservative problem of the SBM with a quasicontinuous spectrum to the LZ nonconservative TLS as described by Eq. \ref{eq_LZ}. In completing the mapping, the time derivative yields the mentioned velocity factor $\dot{q}_{C}$ in the denominator,

\begin{equation}
P_{\uparrow \uparrow }(t\rightarrow \infty )=\exp \left[ -\frac{2\pi }{\hbar
}\frac{\left\vert V_{\uparrow \downarrow }\right\vert ^{2}}{2m\omega
_{0}^{2}q_{C}\dot{q}_{C}}\right] .  \label{eq_PLZ}
\end{equation}
Here, the velocity $\dot{q}_{C}$ can be estimated with a classical calculation of energy conservation: $\dot{q}_{C}=\sqrt{2/m\left[ E_{0}-E_{C}\right] }$.

In Fig. \ref{fig_probLZ} we compare the theoretical value of the LZ probability, given by Eq.\ref{eq_PLZ}, with the numerical value obtained by the evaluation of the first drop of the LE. The comparison is performed as a function of $V_{\uparrow \downarrow }$ for three different values of energy (and hence three different velocities at the crossing). The excellent agreement between them implies the accuracy of the LZ physical picture. This is quite remarkable since the LZ formula gives an asymptotic transition probability and relies on the linear approximation for the energies described above. As discussed in the literature \cite{LZ_review}, the exact dynamics through a nonlinear crossing might evidence transient oscillations that are not described by the LZ formula. These fluctuations occur within a time scale \cite{mullenPRL89,*Vitanov} given, in the sudden limit, by $\tau^{}_Z\approx \sqrt{\hbar} \left[\lim_{V_{AB}\rightarrow 0}\frac{d}{dt}\left( \epsilon _{\uparrow }-\epsilon _{\downarrow }\right)\right]^{-1/2}$, where $\epsilon _{\uparrow }$ and $\epsilon _{\downarrow }$ are the actual  instantaneous eigenenergies, without any linear approximation. In our case, fluctuations at such a small scale $\tau^{}_Z\omega^{}_0\approx 0.05$ would manifest as noise. However, this effect is not observed in the well-defined first step of Fig. \ref{fig LE1}. Thus, the linear LZ provides a good description of this first transition.

Notice that, at a given size of the gap ($V_{\uparrow \downarrow }$ fixed), the transition probability is greater\ when the speed at the crossing is higher. For $V_{\uparrow \downarrow }>5\hbar \omega _{0}$ we observe a strong decoherence, i.e., almost half of the wave packet flips its spin projection. As a consequence, $\mathcal{S}$ will rapidly reach a mixed state irrespective of the initial state being a cat or an incoherent superposition. This may hinder the relative \textit{fragility} of these states. For such reason we consider a $V_{\uparrow \downarrow }$ $\gtrsim\hbar \omega _{0}$ which is still the nonperturbative regime.

Just before the third step a revival shows up. Such revivals repeat in every following step, as can be seen in Fig. \ref{fig LE1}. This can be understood by the semiclassical picture of Fig. \ref{fig_model}(b) since the LE peak appears exactly at the crossing time between the original wave packet and the one that escaped to the second parabola. In other words, successive passages yield substantial interferences between the $\left\vert \uparrow \right\rangle $ and $\left\vert \downarrow \right\rangle $\
components of the evolved wave packet. Such particular interferences yielding the revivals is in fact a manifestation of the well-known St\"{u}ckelberg phase commensuration \cite{LZ_review}, which appears in a TLS when a periodic driving force leads to consecutive passages through an avoided
crossing.

It is also notable that the successive LE steps become deeper as the phase coherence within each wave packet begins to decay. In this regime, the single-passage formula \ref{eq_PLZ} is no longer expected to be valid. However, under the rough assumption that every time the wave packet goes through the avoided crossing a\ LZ process occurs, i.e., Eq. \ref{eq_PLZ}, one can compute a characteristic decay time $\tau _{\phi }$ in a Markovian approximation. This would be given by the fraction of HO cycles needed to reach a specific decay,

\begin{equation}
e^{-1}=(P_{LZ})^{2n},
\end{equation}
since for every cycle the wave packet goes two times through the gap. Then,

\begin{equation}
1/n=-2\ln [P_{LZ}^{{}}]=\frac{2\pi }{\hbar }\frac{\left\vert V_{\uparrow
\downarrow }\right\vert ^{2}}{m\omega _{0}^{2}q_{C}^{{}}\dot{q}_{C}^{{}}}.
\end{equation}

Since the period of oscillation is constant, $\tau _{\phi }=n\left( 2\pi
/\omega _{0}\right) $:

\begin{eqnarray}
\frac{1}{\tau _{\phi }} &=&\frac{2\pi }{\hbar }\left\vert V_{\uparrow
\downarrow }\right\vert ^{2}\frac{1}{2\pi m\omega _{0}^{{}}q_{C}^{{}}\dot{q}%
_{C}^{{}}}  \label{eq_exponential1} \\
&=&\frac{2\pi }{\hbar }\left\vert V_{\uparrow \downarrow }\right\vert ^{2}%
\left[ \frac{1}{4\pi \sqrt{E_{C}\left( E_{0}-E_{C}\right) }}\right]
\label{eq_exponencial2} \\
&=&\left\vert \frac{V_{\uparrow \downarrow }}{\hbar }\right\vert ^{2}2\pi
\hbar N_{1}(E_{0})=\left( \tau _{\uparrow \downarrow }^{{}}\right) ^{-2}\tau
_{\downarrow },
\end{eqnarray}
where $E_{0}$ is the energy of the initial state and $E_{C}$ is the gap energy,
which is given by Eq.\ref{eq_cross}. Therefore, the decay rate diverges as
the difference between $E_{0}$ and $E_{C}$ vanishes. Additionally, initial
states with high energies $E_{0}$ have lower decay. The last line describes
the Markovian decay rate in terms of the density of directly connected
states $N_{1}(E_{0}),$ and in terms of the characteristic time scales $\tau
_{\uparrow \downarrow }^{{}}=\hbar /\left\vert V_{\uparrow \downarrow
}\right\vert $ and $\tau _{\downarrow }=2\pi \hbar N_{1}(E_{0})$.

The corresponding exponential decay may be seen as a Markovian approximation
to the LE degradation with respect to the spin-flip process. As shown in Fig. \ref%
{fig LE1}, the comparison with the actual LE decay is only valid during the
first cycle. Repeated passages would give rise to memory effects which are
not contained in a successive application of the single-passage LZ formula.
Quite remarkably, we observe that the LE turns out to be well fitted by a
Gaussian $M(t)=\exp [-\frac{1}{2}(t/\tau _{G})^{2}]$ . Within a considerably
large energy range, the observed Gaussian time scale turns out to be about 1/3
of the Markovian time, i.e. $\tau _{G}\simeq \tau _{\phi }/3$.

In the context of a spin system interacting with a spin bath, Zurek and
coworkers \cite{zurekpolonica} have argued that a Gaussian decay of a LE can
be identified with a random walk in the energy space. In the Fock-space
representation of our system, it is clear that decoherence is a concatenated
process: the spin flip controlled by $\tau _{\uparrow \downarrow
}^{{}}=\hbar /\left\vert V_{\uparrow \downarrow }\right\vert $ followed by
quantum diffusion along the energy coordinate [vertical chain in Fig. \ref%
{fig_model} (a)]. This last can be identified with such quantum random walk,
with the survival probability given by $\left\vert J_{0}(2t\sqrt{n}V_{g}/\hbar
)\right\vert ^{2}$. For short times this survival turns out to be a condition to
maintain coherence between both spin states, and it is essentially a
Gaussian with a time scale $\tau _{\downarrow QD}(E_{0})=\frac{\mathbf{4}}{%
\hbar }\sqrt{E_{0}E_{C}}$. Thus, it interesting to note that in spite of a
numerical factor, the non-Markovian Gaussian decoherence rate is still
described by Eq.\ref{eq_exponencial2}. This feature is also present in the
Gaussian to exponential interpolation formula proposed by Flambaum and
Izrailev \cite{flambaum2000aust,Izrailev2001} for short times, when memory
effects are still effective. However, in spite of the mentioned plausibility
arguments they are more appropriate to describe the decay of single-energy
eigenstates, but they are not enough to provide a quantitative description of
the degradation of the subtle collective interferences involved in the
semiclassical wave-packet dynamics.

\subsection{Decoherence and \textit{fragility}}

\label{sub_fragility}

\begin{figure*}
    \centering
     \includegraphics[width=0.95\textwidth]{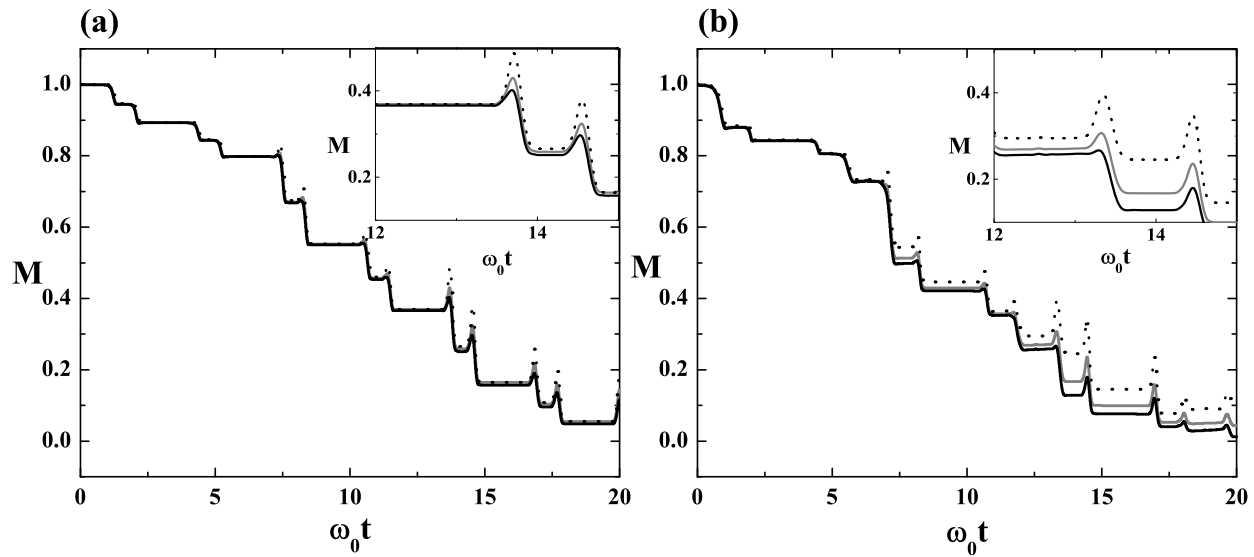}\\
      \caption{Comparison between the LE for a cat state (black line), the LE for an incoherent superposition (gray line)
as given by Eq.\ref{app_LEincoh2} and the average LE of two independent evolutions of semiclassical states (dotted line) as given by Eq.\ref{LE_Naive}. The LE dynamics is given by $\overline{E}=150\hbar \omega _{0}$, $E_{\uparrow }=0$, $E_{\downarrow }=100\hbar 
\omega _{0}$, $V_{g}=10\hbar \protect\omega _{0}$, $V_{\uparrow \downarrow }=2\hbar\omega _{0}$. (a) $\Delta E=0$ and (b) $\Delta E=200\hbar \protect\omega _{0}$.}%
    \label{fig_le2}%
\end{figure*}

In order to study the \textit{fragility} of the cat state defined in Eq.\ref{eq cat}, we fix its mean energy $\overline{E}\equiv (E_{1}+E_{2})/2=\left(\alpha _{1}^{2}+\alpha _{2}^{2}+1\right) /2\gg E_{C}$ and analyze the LE for a set of energy differences $\Delta E\equiv \left\vert E_{1}-E_{2}\right\vert =\left\vert \alpha _{1}^{2}-\alpha _{2}^{2}\right\vert $. Here, $E_{1}$ and $E_{2}$ are the energies of all individual semiclassical wave packets that compose the cat state. We summarize the observed behavior by plotting $M(t)$ for two representative cases in Fig. \ref{fig_le2}. There, we compare the LE for an initial cat state, an initial incoherent superposition, and an average value of the LEs corresponding to the independent dynamics of the two individual semiclassical states [see Eq.\ref{LE_Naive} in Appendix]. When $\Delta E=0$ [Fig.\ref{fig_le2}(a)], there is almost no difference between the behavior
of the LE for the three cases since the steps show up at the same time and have the same depth. We assign this effect to a particularity of the interaction used which, being energy dependent, produces an equivalent change in quantum phases of each wave packet of the superposition. Thus, only the adiabatic tunneling would contribute to the decoherent process, and this has the same effect for the cat state and for the incoherent superposition. The difference between the two cases relies on the revivals associated with the St\"{u}ckelberg phase, which occurs when the wave packet components that remained with the same spin state interfere with the ones that changed it. In fact, the incoherent superposition state shows larger revivals. When $\Delta E\neq 0$ the situation changes. For $\overline{E}=150\hbar \omega _{0}$ the cat state degrades faster than the incoherent superposition as $\Delta E$ is increased; i.e., the LE for the cat state tends to be lower. This means that as $\Delta E$ increases, the nonlocal (in energy) states become more fragile [Fig. \ref{fig_le2} (b)]. The nonlocality in space is not sufficient to ensure a difference in the behavior of different initial states. In particular, if $\alpha _{1}=\left\vert \alpha \right\vert $ and $\alpha _{2}=-\left\vert \alpha \right\vert$, then $\Delta E=0$ but $\Delta q\neq 0$. As shown in Fig. \ref{fig_le2}(a), this case does not show evidence of relative fragility.

\begin{figure}
    \centering
     \includegraphics[width=0.5\textwidth]{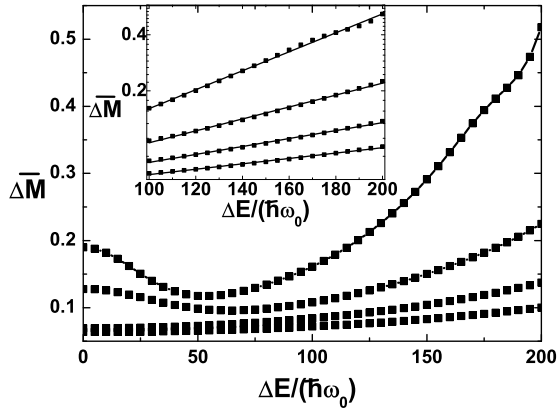}\\
      \caption{$\Delta \bar{M}$ as function of $\Delta E$. From top to bottom, $\overline{E}=150\hbar \protect\omega _{0}$,
$\overline{E}=200\hbar \protect\omega _{0}$, $\overline{E}=250\hbar \omega _{0}$ and $\overline{E}=300\hbar \protect\omega _{0}$. The inset shows $\Delta \bar{M}$ as a function of $\Delta E$ in log scale. The linear fittings indicate an asymptotic exponential dependence for $\Delta E\gtrsim 100\hbar \protect\omega _{0}$.}
    \label{fig area}%
\end{figure}

In order to better quantify the previous observations, we define the mean LE as $\bar{M}=1/T\int_{0}^{T}M(t)dt$. At $T=20/\omega _{0}$ we compute the difference $\Delta \bar{M}=$ $\left\vert \bar{M}_{inc}-\bar{M}_{cat}\right\vert $, where $\bar{M}_{inc}$ and $\bar{M}_{cat}$ indicate the mean LE of the incoherent superposition and the cat state respectively. Thus, $\Delta \bar{M}$ corresponds to the area between the two curves, $M_{inc}$ and $M_{cat}$. The magnitude $\Delta \bar{M}$ constitutes our \textit{fragility }quantifier. In Fig. \ref{fig area} we show how $\Delta\bar{M}$ increases with $\Delta E$ for different $\overline{E}$ provided that $\Delta E\gtrsim 100\hbar \omega _{0}$. The scaling law turns out to be exponential on the energy difference: $\Delta \bar{M}\sim \exp [(\Delta E)/\overline{E}^{\nu }]$, with $\nu \simeq 3.5$. This means that the fragility of the cat state increases as the nonlocality in energy grows. Also, it can be noticed that $\Delta \bar{M}$ does not vanish even at $\Delta E=0$. Two observations contribute to the interpretation of such an effect. On the one hand $\Delta E$ must exceed the natural energy uncertainty of each of the individual wave packets forming the initial state. On the other hand, since many LZ processes contribute to $\Delta \bar{M},$ a finite value for $\Delta E=0$ can be associated with the revivals that appear immediately before LZ processes that define the LE steps. As they originate in a precise phase commensuration, the more fragile cat states always have smaller revivals than those of the incoherent superposition of wave packets. Such an effect is more noticeable when $\overline{E}$\ gets closer to $E_{C}$. Additionally, Fig. \ref{fig area} shows that the fragility tends to disappear and the effects of $\Delta E$ diminish as $\overline{E}$\ increases. Indeed, if $\overline{E}$ is very high, the LE for the cat state does not present changes in its behavior even when $\Delta E$ varies for a wide range of values. We can interpret this fact as a consequence of the energy dependence of the perturbation, which for initial states with large $\overline{E}$ implies a LZ\ factor of almost $1$. Thus, the tunneling through the avoided crossing is negligible, and the perturbation is less effective as a decoherent process.

\section{Conclusions}

\label{sec_conclusions}

In this article we employed the spin-boson model to study decoherence of a harmonic oscillator produced by its interaction with a simple nondegenerate binary environment. Since here the system has many more allowed states than the environment, the conceptual approach contrasts the standard picture of quantum open systems. A particularity of our model is that the spin-flip process only becomes effective when the oscillator coordinate is such that the interaction energy makes both spin states degenerate. Thus, spin-flip dynamics within the environment is limited to occur only at a precise coordinate of the harmonic oscillator. It is quite remarkable that this situation, involving an unbounded set of discrete energies, turns out to be well described in terms of the Landau-Zener theory, which was developed for a two level system evolving under a time dependent energy splitting.

The degradation of the quantum phase produced by the environment was quantified by the Loschmidt echo. In particular, we focused on the fragility of the coherent superposition of wave packets (cat states) when compared with incoherent superpositions of the same wave packets. This required the evaluation of the dynamics of such states. A tool that made such calculations even more handleable was a wave-function treatment, which involves a chosen number $N$ of states in the Hilbert space, instead of a full density matrix, which would involve dimensions of $N\times N$. The results indicate that coherent superpositions of semiclassical wave packets associated with different energies are more fragile than incoherent ones. The fragility increases with the energy difference, i.e., nonlocality in the energy representation, between the individual wave packets. In our model, there is no evidence of fragility strictly related to spatial nonlocality. However, when nonlocality in space is associated with nonlocality in energy, the system becomes increasingly fragile towards the simple decoherence process.

The exponentially increased fragility of cat states may be related to the problem of thermalization in closed quantum systems \cite{PolkovnikovRevModPhys}. In Ref. \cite{deutsch} it was proposed to build a cat state with two macroscopic (semiclassical) wave functions with different energies for the purpose of analyzing the time average of any particular observable. In such a situation, in order to recover the standard (classical) microcanonical predictions for such observables, the interferences between the wave functions should be negligible. Our results constitute a step towards this direction since we verified that the more separated in energy these wave packets are, the more easily they decorrelate, i.e., the more fragile its phase coherence becomes.

\section{Acknowledgments}

We acknowledge financial support from CONICET, ANPCyT, SeCyT-UNC and MinCyT-Cor. This work benefited from discussions with A. D. Dente and L. J. Fern\'{a}ndez-Alc\'{a}zar. D.B. kindly acknowledges M. Castagnino for introducing her to the decoherence problem and his constant support during the first stages of this work. 

\begin{widetext}

\appendix

\section{Incoherent States}

\subsection{Normalization}

We summarize here the properties of the incoherent superposition of two Gaussians given by Eq.\ref{eq incoh},
\begin{equation}
\left\vert \Psi _{0}^{inc}\right\rangle =\lim_{N\rightarrow \infty }\sum\limits_{j=1}^{N}\frac{e^{i\theta _{j}}\left\vert \alpha_{1}\right\rangle +e^{i\phi _{j}}\left\vert \alpha_{2}\right\rangle }{\sqrt{\Delta }},  \label{app_incoh1}
\end{equation}
where the normalization is
\begin{equation}
\Delta ^{2}=2N^{2}\left[ 3+\exp \left( -\left\vert \alpha _{1}-\alpha
_{2}\right\vert ^{2}\right) \right] .  \label{app_incoh2}
\end{equation}%
To find this value we employed an algebra that is useful also for the calculation of the LE. It follows from the use of the identities, which hold for $N$ sufficiently large,
\begin{equation}
\sum\limits_{j=1}^{N}\exp [i\theta_{j}]=\sum\limits_{j=1}^{N}\exp [i\phi _{j}]=0,  \label{aux1}
\end{equation}
\begin{equation}
\sum\limits_{j,j^{\prime }}^{N}\exp [i(\theta _{j}-\theta_{j^{\prime }})]=N+\sum\limits_{j\neq j^{\prime }}^{N}\exp [i
(\theta _{j}-\theta _{j^{\prime }})]=N,  \label{aux2}
\end{equation}
\begin{equation}
\sum\limits_{j,j^{\prime }}^{N}\exp \left[ i\left( \phi _{j}-\phi _{j^{\prime }}\right) \right] =N+\sum\limits_{j\neq j^{\prime }}^{N}\exp \left[ i\left( \phi _{j}-\phi _{j^{\prime }}\right) \right] =N,
\label{aux3}
\end{equation}
and hence
\begin{equation}
\sum\limits_{j,j^{\prime },s,s^{\prime }}^{N}\exp [i(\theta _{j}-\theta _{j^{\prime }}+\theta _{s^{\prime }}-\phi _{s})]=0,  \label{aux4}
\end{equation}
\begin{equation}
\sum\limits_{j,j^{\prime }}^{N}\exp [\pm i(\theta _{j}+\theta _{j^{\prime }})]=\sum\limits_{s,s^{\prime }}^{N}\exp [\pm i(\phi
_{s^{\prime }}+\phi _{s})]=0,  \label{aux4bis}
\end{equation}
\begin{align}
\sum\limits_{j,j^{\prime },s,s^{\prime }}^{N} & \exp [i(\theta _{j}-\theta _{j^{\prime }}+\phi _{s^{\prime }}-\phi
_{s})] \nonumber \\
 &= \sum\limits_{j=j^{\prime },s=s^{\prime }}^{N}\exp [i(\theta
_{j}-\theta _{j^{\prime }}+\phi _{s^{\prime }}-\phi _{s})]=N^{2},
\label{aux5}
\end{align}

\begin{align}
\sum\limits_{j,j^{\prime },s,s^{\prime }}^{N} & \exp [i(\theta
_{j}-\theta _{j^{\prime }}+\theta _{s^{\prime }}-\theta _{s})]  \nonumber \\
&= \left[\sum\limits_{j=j^{\prime }\neq s=s^{\prime }}^{N}+\sum\limits_{j=s\neq
j^{\prime }=s^{\prime }}^{N}+\sum\limits_{j=j^{\prime }=s=s^{\prime }}^{N}
\right]  \nonumber \\
& \times  \exp [i(\theta _{j}-\theta _{j^{\prime }}+\theta _{s^{\prime }}-\theta _{s})] \nonumber \\
& = N(N-1)+N(N-1)+N=2N^{2}-N\simeq 2N^{2}.
\label{aux6}
\end{align}

Normalization can be then computed by writing
\begin{eqnarray}
\left\vert \left\langle \Psi _{0}^{inc}|\Psi _{0}^{inc}\right\rangle
\right\vert ^{2} &=&\Delta ^{-2}\sum\limits_{j,j^{\prime },s,s^{\prime
}}^{N}\left\vert \langle \alpha _{1}\left\vert \alpha _{1}\right\rangle
\right\vert ^{2}\exp [i\left( \theta _{j}-\theta _{j^{\prime }}+\theta _{s^{\prime }}-\theta _{s}\right) ]+\langle \alpha _{1}\left\vert \alpha _{1}\right\rangle \langle \alpha _{1}\left\vert \alpha _{2}\right\rangle \exp [i(\theta _{j}-\theta _{j^{\prime }}+\phi_{s^{\prime }}-\theta _{s})]  \nonumber \\
&&+\langle \alpha _{1}\left\vert \alpha _{1}\right\rangle \langle \alpha
_{2}\left\vert \alpha _{1}\right\rangle \exp [i(\theta _{j}-\theta
_{j^{\prime }}+\theta _{s^{\prime }}-\phi _{s})]+\langle \alpha
_{1}\left\vert \alpha _{1}\right\rangle \langle \alpha _{2}\left\vert \alpha
_{2}\right\rangle \exp [i(\theta _{j}-\theta _{j^{\prime }}+\phi
_{s^{\prime }}-\phi _{s})]  \nonumber \\
&&+\langle \alpha _{1}\left\vert \alpha _{2}\right\rangle \langle \alpha
_{1}\left\vert \alpha _{1}\right\rangle \exp [i(\phi _{j}-\theta
_{j^{\prime }}+\theta _{s^{\prime }}-\theta _{s})]+\left\vert \langle \alpha
_{1}\left\vert \alpha _{2}\right\rangle \right\vert ^{2}\exp [i(\phi _{j}-\theta _{j^{\prime }}+\theta _{s^{\prime }}-\phi _{s})]
\nonumber \\
&&+\langle \alpha _{1}\left\vert \alpha _{2}\right\rangle ^{2}\exp [i(\phi _{j}-\theta _{j^{\prime }}+\phi _{s^{\prime }}-\theta _{s})]+\langle\alpha _{1}\left\vert \alpha _{2}\right\rangle \langle \alpha _{2}\left\vert
\alpha _{2}\right\rangle \exp [i(\phi _{j}-\theta _{j^{\prime }}+\phi _{s^{\prime }}-\phi _{s})]  \nonumber \\
&&+\langle \alpha _{2}\left\vert \alpha _{1}\right\rangle \langle \alpha _{1}\left\vert \alpha _{1}\right\rangle \exp [i(\theta _{j}-\phi _{j^{\prime }}+\theta _{s^{\prime }}-\theta _{s})]+\langle \alpha _{2}\left\vert \alpha _{1}\right\rangle ^{2}\exp [i(\theta _{j}-\phi _{j^{\prime }}+\theta _{s^{\prime }}-\phi _{s})]  \nonumber \\
&&+\left\vert \langle \alpha _{2}\left\vert \alpha _{1}\right\rangle
\right\vert ^{2}\exp [i(\theta _{j}-\phi _{j^{\prime }}+\phi
_{s^{\prime }}-\theta _{s})]+\langle \alpha _{2}\left\vert \alpha
_{1}\right\rangle \langle \alpha _{2}\left\vert \alpha _{2}\right\rangle
\exp [i(\theta _{j}-\phi _{j^{\prime }}+\phi _{s^{\prime }}-\phi
_{s})]  \nonumber \\
&&+\langle \alpha _{2}\left\vert \alpha _{2}\right\rangle \langle \alpha
_{1}\left\vert \alpha _{1}\right\rangle \exp [i(\phi _{j}-\phi
_{j^{\prime }}+\theta _{s^{\prime }}-\theta _{s})]+\langle \alpha
_{2}\left\vert \alpha _{2}\right\rangle \langle \alpha _{2}\left\vert \alpha
_{1}\right\rangle \exp [i(\phi _{j}-\phi _{j^{\prime }}+\theta
_{s^{\prime }}-\phi _{s})]  \nonumber \\
&&+\langle \alpha _{2}\left\vert \alpha _{2}\right\rangle \langle \alpha
_{1}\left\vert \alpha _{2}\right\rangle \exp [i(\phi _{j}-\phi
_{j^{\prime }}+\phi _{s^{\prime }}-\theta _{s})]+\left\vert \langle \alpha
_{2}\left\vert \alpha _{2}\right\rangle \right\vert ^{2}\exp [i%
(\phi _{j}-\phi _{j^{\prime }}+\phi _{s^{\prime }}-\phi _{s})]
\label{normaliza1}
\end{eqnarray}%

and noticing that $\langle \alpha _{1}\left\vert \alpha _{1}\right\rangle
=\langle \alpha _{2}\left\vert \alpha _{2}\right\rangle =1$ and $\left\vert
\langle \alpha _{1}\left\vert \alpha _{2}\right\rangle \right\vert
^{2}=\left\vert \langle \alpha _{2}\left\vert \alpha _{1}\right\rangle
\right\vert ^{2}=\exp [-\left\vert \alpha _{1}-\alpha _{2}\right\vert ^{2}]$. By Eq.\ref{aux4}, the $2^{nd}$, $3^{rd}$, $5^{th}$, $8^{th}$, $9^{th}$, $12^{th}$, $14^{th}$, and $15^{th}$ terms vanish, and by Eq.\ref{aux4bis}
the $7^{th}$ and $10^{th}$ terms also vanish. Using Eqs.\ref{aux5} and \ref{aux6}, normalization in Eq.\ref{normaliza1} finally yields
\begin{align*}
& \left\vert \left\langle \Psi _{0}^{inc}|\Psi _{0}^{inc}\right\rangle
\right\vert ^{2}  \\
&= \Delta ^{-2}\left[ 2N^{2}+2N^{2}+2N^{2}\exp [-\left\vert
\alpha _{1}-\alpha _{2}\right\vert ^{2}]+2N^{2}\right]  \\
&=\Delta ^{-2}2N^{2}\left( 3+\exp [-\left\vert \alpha _{1}-\alpha
_{2}\right\vert ^{2}]\right),  \\
&=1,
\end{align*}%
which defines $\Delta $ as in Eq.\ref{app_incoh2}.

\subsection{Time evolution and LE}

In order to compute the LE as defined in Eq.\ref{eq_LEformula2}, one shall consider two different cases. For single semiclassical states as in Eq.\ref{eq coherent} or cat superpositions as in Eq.\ref{eq cat}, the LE can be straightforwardly evaluated by Eq.\ref{eq_LEformula3}. However, for the incoherent superpositions of Eq.\ref{eq incoh}, an appropriate manipulation is required. Thus, here we make explicit the time evolution and the LE for such state. With the purpose of simplifying notation, $\lim_{N\rightarrow \infty }$ is dropped everywhere.

First, notice that in the Fock basis $\left\vert \Psi
_{0}^{inc}\right\rangle $ can be written in a split form making explicit
the random phases $\Lambda $ and the amplitudes $c_{k,n}^{{}}$ needed to
build each of the Gaussian coherent states:

\begin{equation}
\left\vert \Psi _{0}^{inc}\right\rangle =\sum\limits_{n=0}^{\infty }\left[
\Lambda _{{}}^{(1)}c_{\uparrow ,n}^{(1)}+\Lambda _{{}}^{(2)}c_{\uparrow
,n}^{(2)}\right] \left\vert \uparrow ,n\right\rangle ,  \label{app_incoh3}
\end{equation}%
where

\begin{eqnarray}
\Lambda _{{}}^{(1)} &=&\sum\limits_{j=1}^{N}\frac{e^{i\theta _{j}}%
}{\sqrt{\Delta }},  \nonumber \\
\Lambda _{{}}^{(2)} &=&\sum\limits_{j=1}^{N}\frac{e^{i\phi _{j}}}{%
\sqrt{\Delta }},  \label{app_fases} \\
c_{\uparrow ,n}^{(1)} &=&\exp [-\frac{\left\vert \alpha _{1}\right\vert ^{2}%
}{2}]\frac{(\alpha _{1})^{n}}{\sqrt{n!}},  \nonumber \\
c_{\uparrow ,n}^{(2)} &=&\exp [-\frac{\left\vert \alpha _{2}\right\vert ^{2}%
}{2}]\frac{(\alpha _{2})^{n}}{\sqrt{n!}}.  \label{app_coef1}
\end{eqnarray}%
Since any evolution operator is linear, the splitting of the probability
amplitudes remains valid at any time. In fact, the evolution under the
Hamiltonian $\hat{H}_{\mathcal{S}}$ can be exactly computed as

\begin{equation}
e^{-\frac{i}{\hbar }\hat{H}_{\mathcal{S}}t}\left\vert \Psi
_{0}^{inc}\right\rangle =\sum\limits_{n=0}^{\infty }\left[ \Lambda
_{{}}^{(1)}c_{\uparrow ,n}^{(1)}(t)+\Lambda _{{}}^{(2)}c_{\uparrow
,n}^{(2)}(t)\right] \left\vert \uparrow ,n\right\rangle ,  \label{app_incoh4}
\end{equation}%
where

\begin{eqnarray}
c_{\uparrow ,n}^{(1)}(t) &=&\exp [-\frac{\left\vert \alpha _{1}\right\vert
^{2}}{2}-i(n+\frac{1}{2})\omega _{0}t]\frac{(\alpha _{1})^{n}}{%
\sqrt{n!}},  \nonumber \\
c_{\uparrow ,n}^{(2)}(t) &=&\exp [-\frac{\left\vert \alpha _{2}\right\vert
^{2}}{2}-i(n+\frac{1}{2})\omega _{0}t]\frac{(\alpha _{2})^{n}}{%
\sqrt{n!}}.  \label{app_coef3}
\end{eqnarray}%
Analogously, the perturbed evolution under $\hat{H}_{\mathcal{S}}+\hat{\Sigma%
}$ yields

\begin{equation}
e^{-\frac{i}{\hbar }(\hat{H}_{\mathcal{S}}+\hat{\Sigma}%
)t}\left\vert \Psi _{0}^{inc}\right\rangle =\sum\limits_{n=0}^{\infty }%
\left[ \Lambda _{{}}^{(1)}c_{\uparrow ,n}^{(1)}(0)+\Lambda
_{{}}^{(2)}c_{\uparrow ,n}^{(2)}(0)\right] e^{-\frac{i}{\hbar }(%
\hat{H}_{\mathcal{S}}+\hat{\Sigma})t}\left\vert \uparrow ,n\right\rangle
\equiv \sum_{n=0}^{\infty }\sum_{k=\uparrow ,\downarrow }\left[ \Lambda
_{{}}^{(1)}d_{k,n}^{(1)}(t)+\Lambda _{{}}^{(2)}d_{k,n}^{(2)}(t)\right]
\left\vert k,n\right\rangle .  \label{app_incoh5}
\end{equation}%

Even though we do not have a simple closed formula like Eq.\ref{app_coef3} for the time-dependent amplitudes $d_{k,n}^{(1)}(t)$ and $d_{k,n}^{(2)}(t)$, they are well defined by the linearity of the evolution operator. Now we can translate the LE evaluation in Eq.\ref{eq_LEformula3} by identifying $c_{\uparrow n}(t)\rightarrow \Lambda _{{}}^{(1)}c_{\uparrow
,n}^{(1)}(t)+\Lambda _{{}}^{(2)}c_{\uparrow ,n}^{(2)}(t)$ and $d_{k,n}^{{}}(t)\rightarrow \Lambda _{{}}^{(1)}d_{k,n}^{(1)}(t)+\Lambda
_{{}}^{(2)}d_{k,n}^{(2)}(t)$,

\begin{align}
M_{inc}(t) &= \text{Tr}\left\{ \left( \sigma _{\mathcal{S}}^{r}\right) ^{\dag
}\sigma _{\mathcal{S+E}}^{r}\right\}   \nonumber \\
& = \sum_{m,n=0}^{\infty } \left[   \left( \Lambda _{{}}^{(1)}d_{\uparrow
,m}^{(1)}(t)+\Lambda _{{}}^{(2)}d_{\uparrow ,m}^{(2)}(t)\right) \left( \bar{%
\Lambda}_{{}}^{(1)}\bar{d}_{\uparrow ,n}^{(1)}(t)+\bar{\Lambda}_{{}}^{(2)}%
\bar{d}_{\uparrow ,n}^{(2)}(t)\right) \right. \nonumber \\
&\left.  \qquad \qquad + \left( \Lambda
_{{}}^{(1)}d_{\downarrow ,m}^{(1)}(t)+\Lambda _{{}}^{(2)}d_{\downarrow
,m}^{(2)}(t)\right) \left( \bar{\Lambda}_{{}}^{(1)}\bar{d}_{\downarrow
,n}^{(1)}(t)+\bar{\Lambda}_{{}}^{(2)}\bar{d}_{\downarrow ,n}^{(2)}(t)\right) %
 \right]   \nonumber \\
& \qquad \qquad \qquad \qquad \times \left[ \left( \Lambda_{{}}^{(1)}c_{\uparrow ,n}^{(1)}(t)+\Lambda
_{{}}^{(2)}c_{\uparrow ,n}^{(2)}(t)\right) \left( \bar{\Lambda}_{{}}^{(1)}%
\bar{c}_{\uparrow ,m}^{(1)}(t)+\bar{\Lambda}_{{}}^{(2)}\bar{c}_{\uparrow
,m}^{(2)}(t)\right) \right],  \label{app_LEincoh1}
\end{align}%
where the overline means complex conjugation. Eq. \ref{app_LEincoh1} has $32$ terms, and after using the averaging rules given by Eqs.\ref{aux4},\ref{aux4bis},\ref{aux5} and \ref{aux6} for every product $\Lambda_{{}}^{(a)}\times \bar{\Lambda}_{{}}^{(b)}\times \Lambda _{{}}^{(c)}\times \bar{\Lambda}_{{}}^{(d)}$,

\begin{align}
M_{inc}(t)  =& \Delta ^{-2}\sum_{m,n=0}^{\infty } \left[  2N^{2}\left( d_{\uparrow
,m}^{(1)}(t)\bar{d}_{\uparrow ,n}^{(1)}(t)+d_{\downarrow ,m}^{(1)}(t)\bar{d}%
_{\downarrow ,n}^{(1)}(t)\right) c_{\uparrow ,n}^{(1)}(t)\bar{c}_{\uparrow
,m}^{(1)}(t) \right.  \nonumber \\
& + N^{2}\left( d_{\uparrow ,m}^{(1)}(t)\bar{d}_{\uparrow
,n}^{(1)}(t)+d_{\downarrow ,m}^{(1)}(t)\bar{d}_{\downarrow ,n}^{(1)}(t)\right)
c_{\uparrow ,n}^{(2)}(t)\bar{c}_{\uparrow ,m}^{(2)}(t)+
N^{2}\left( d_{\uparrow ,m}^{(2)}(t)\bar{d}_{\uparrow
,n}^{(1)}(t)+d_{\downarrow ,m}^{(2)}(t)\bar{d}_{\downarrow
,n}^{(1)}(t)\right) \bar{c}_{\uparrow ,m}^{(2)}(t)c_{\uparrow
,n}^{(1)}(t) \nonumber \\
& + N^{2}\left( d_{\uparrow ,m}^{(1)}(t)\bar{d}_{\uparrow
,n}^{(2)}(t)+d_{\downarrow ,m}^{(1)}(t)\bar{d}_{\downarrow
,n}^{(2)}(t)\right) \bar{c}_{\uparrow ,m}^{(1)}(t)c_{\uparrow ,n}^{(2)}(t)
+N^{2}\left( d_{\uparrow ,m}^{(2)}(t)\bar{d}_{\uparrow
,n}^{(2)}(t)+d_{\downarrow ,m}^{(2)}(t)\bar{d}_{\downarrow
,n}^{(2)}(t)\right) \bar{c}_{\uparrow ,m}^{(1)}(t)c_{\uparrow
,n}^{(1)}(t) \nonumber \\
&\left. +2N^{2}\left( d_{\uparrow ,m}^{(2)}(t)\bar{d}_{\uparrow
,n}^{(2)}(t)+d_{\downarrow ,m}^{(2)}(t)\bar{d}_{\downarrow
,n}^{(2)}(t)\right) \bar{c}_{\uparrow ,m}^{(2)}(t)c_{\uparrow ,n}^{(2)}(t)\right] .
\label{app_LEincoh2}
\end{align}%

Notice that since $\Delta ^{-2}\propto N^{-2}$, the $N$ dependence of Eq.\ref{app_LEincoh2} disappears. This means that we do not
need to compute an infinite average of wave functions. Instead, it is only needed to evolve separately two individual semiclassical states $\left\vert \alpha _{1}\right\rangle $ and $\left\vert \alpha _{2}\right\rangle $ and use their respective probability amplitudes (the complex coefficients in the Fock basis $\left\{ \left\vert k,n\right\rangle \right\} $) to compute $M_{inc}(t)$ at any time. Additionally, the first and last terms in Eq.\ref{app_LEincoh2} are proportional to the naive version of the LE for two independent semiclassical states, defined as the mean value of the individual overlaps:

\begin{eqnarray}
M_{naive}(t) &\equiv &\frac{1}{2}\sum_{m,n=0}^{\infty }\left[ \left(
d_{\uparrow ,m}^{(1)}(t)\bar{d}_{\uparrow ,n}^{(1)}(t)+d_{\downarrow
,m}^{(1)}(t)\bar{d}_{\downarrow ,n}^{(1)}(t)\right) c_{\uparrow ,n}^{(1)}(t)%
\bar{c}_{\uparrow ,m}^{(1)}(t)+\left( d_{\uparrow ,m}^{(2)}(t)\bar{d}%
_{\uparrow ,n}^{(2)}(t)+d_{\downarrow ,m}^{(2)}(t)\bar{d}_{\downarrow
,n}^{(2)}(t)\right) \bar{c}_{\uparrow ,m}^{(2)}(t)c_{\uparrow ,n}^{(2)}(t)%
\right]  \nonumber \\
&=&\frac{1}{2}\left( M_{{}}^{(\alpha _{1})}(t)+M_{{}}^{(\alpha
_{2})}(t)\right) ,  \label{LE_Naive}
\end{eqnarray}

where $M_{{}}^{(\alpha _{1})}$ and $M_{{}}^{(\alpha _{2})}$ are the
corresponding LE for $\left\vert \alpha _{1}\right\rangle $ and $\left\vert
\alpha _{2}\right\rangle $ respectively, evaluated from Eq. \ref%
{eq_LEformula3}.
\end{widetext}



\end{document}